\UseRawInputEncoding
\documentclass{article}

\usepackage[utf8]{inputenc}
\usepackage[final]{neurips_2022}
\usepackage[english]{babel}
\usepackage[pdftex]{graphicx}
\usepackage{subcaption}
\usepackage{natbib}
\usepackage[T1]{fontenc}    % use 8-bit T1 fonts
\usepackage{hyperref}       % hyperlinks
\usepackage{url}            % simple URL typesetting
\usepackage{booktabs}       % professional-quality tables
\usepackage{amsfonts}       % blackboard math symbols
\usepackage{nicefrac}       % compact symbols for 1/2, etc.
\usepackage{microtype}      % microtypography
\usepackage{xcolor}         % colors
\usepackage{amsmath}
\DeclareUnicodeCharacter{0394}{\ensuremath{\Delta}}

\title{Understanding stock market instability via graph auto-encoders}

\author{%
  Dragos Gorduza \\
  University of Oxford \\
  Oxford, United Kingdom \\
  \texttt{dragos.gorduza@st-annes.ox.ac.uk} \\
  \AND
  Stephan Zohren \\
  University of Oxford \\
  Oxford, United Kingdom \\
  \texttt{stefan.zohren@eng.ox.ac.uk} \\
  \And
  Xiaowen Dong \\
  University of Oxford \\
  Oxford, United Kingdom \\
  \texttt{xdong@robots.ox.ac.uk} \\
}

\begin{document}

\maketitle

\begin{abstract}
Understanding stock market instability is a key question in financial management as practitioners seek to forecast breakdowns in asset co-movements which expose portfolios to rapid and devastating collapses in value. The structure  of these co-movements can be described as a graph where companies are represented by nodes and edges capture correlations between their price movements. Learning a timely indicator of co-movement breakdowns (manifested as modifications in the graph structure) is central in understanding both financial stability and volatility forecasting. We propose to use the edge reconstruction accuracy of a graph auto-encoder (GAE) as an  indicator for how spatially homogeneous connections between assets are, which, based on financial network literature, we use as a proxy to infer market volatility. Our experiments on the S\&P 500 over the 2015-2022 period show that higher GAE reconstruction error values are  correlated with higher volatility. We also show that out-of-sample autoregressive modeling of volatility is improved by the addition of the proposed measure. 
Our paper contributes to the literature of machine learning in finance particularly in the context of understanding stock market instability.
\end{abstract}
{\bf Keywords}\\ \small{Graph Based Learning, Graph Neural
Networks, Graph Autoencoder, Stock Market Information, Volatility Forecasting}

\section{Introduction}
% catchy title section
Financial market instability has long interested investors, as, during financial collapses, previously uncorrelated companies collapse in value together, which exposes the market to a systemic level of risk that goes beyond the level of the company and requires a more global outlook \citep{Andersen2006}. Understanding this market-wide systemic risk driven by asset co-movement requires framing connections between firms as formed by evolving `complex adaptive networks' \citep{Haldane2009}. This turns instability detection into a network analysis task. 
%Effective portfolio building requires estimating how assets in a market evolve together \citep{Mantegna1999,Kritzman2011}. %This is particularly crucial during financial turmoil when the homogeneity of correlation structures disappears as diversification of risks 'breaks down' when it is most needed, \citep{Preis2012} and all stocks in a portfolio collapse in value together \citep{Sandoval2012,Zhang2020,Poon2003,Campbell2002}.  
These networks are usually constructed by transforming and filtering correlation matrices of price returns \citep{Mantegna1999}. 
They capture key structures of the correlation matrices, and, at the same time they avoid potentially noisy features derived from correlations directly \citep{Avellaneda2020}.
%Networks are useful as forecasting systemic risk directly with correlation matrices yields measures that are unstable, noisy, vulnerable to outliers from high variance time series with power-law distributions as found in finance \citep{Avellaneda2020}. 
Topological features of the resulting  financial graphs, such as shorter diameters \citep{Mantegna1999}, higher cluster density \citep{Mantegna1999,Onnela2004,Tumminello2010}, or lower Ricci curvature \citep{Samal2021,Sandhu2016}, are shown to be positively linked to market instability measures such as higher volatility.% as empirically observed in Fig. \ref{fig:corr-network-vol}. 

This suggests that higher market instability results from breakdowns to previously homogeneous patterns of network connectivity. Homogeneity is defined here as regular connection patterns across the entire network \cite{Xue2022}. In financial networks, homogeneous patterns are connections between companies of the same industry sectors and with similar returns \citep{Kukreti2020}. Homogeneity-disturbing shocks arise from news announcements about financial events or asset bubbles and affect the ways in which the assets co-move by modifying the way investors form opinions about how companies are related and how they will co-evolve \citep{Campbell2002,Lim,Preis2012}.  From this perspective, market instability can be seen as modifying the underlying mechanism through which inter-company connectivity structures appear. \citet{Kukreti2020} points out that heterogeneously connected graphs indicates higher instability, as any shock could travel further across the entire market and not be limited to a single sector \citep{Artzner1999,Das2019}. Heterogeneity in a network is defined as difference over all pairs of linked nodes of a function applied to node features \cite{Estrada2010}. Heterogeneous market networks are therefore characterised by high connectivity between companies which differ in industry sector, market capitalisation or long-run returns.  Increase in this heterogeneity has been shown to granger-cause increases in the VIX volatility index \cite{Zhao2016}.  Conversely, more homogeneous periods when strong connections only exist between similar assets would indicate lower risk, as any shock would propagate at most throughout that asset's immediate neighbourhood in the financial graph and not the entire market \citep{Samal2021,Heiberger2014}.
% Figure 1 being outcommented

%\begin{figure}[!ht]
%    \centering
%    \includegraphics[width=0.8\textwidth
%    %,height = 45mm 
%    ]{images/time_series/more_heterogeneous_more_vol.png}
%    \caption{Market instability is positively linked to connectivity heterogeneity and graph clustering}
%    \label{fig:corr-network-vol}
%\end{figure}

However, the ``network-only'' approaches mentioned above may not fully capture homogeneity of connections between different types of nodes. Indeed, once networks are constructed, they do not make further use of any form of node features. This means they implicitly assume that all information about company nodes is captured by their connectivity. %thus treating all companies in a financial network as identical. 
This discount information from the underlying price returns which traditional volatility forecasting models like multivariate GARCH \citep{Chandra} or HEAVY \citep{Noureldin2012} successfully use in forecasting volatility. Thus, a measure which accounts for both the network and node features is lacking. This leads to the open question of how to measure market instability using financial networks with company characteristics such as returns. 

The graph machine learning literature has provided tools for the joint learning of these features \citep{Chami2022}. However, their applications to financial graphs remain mostly limited to return forecasting, with very few works looking at volatility forecasting directly \citep{Saha2022}. \citet{Wang2020} point to the lack of unsupervised models in graph neural network applications to the stock  market that can learn representations of market states instead of only directly forecasting them. We propose a graph auto-encoder (GAE) \citep{Kipf2016} to capture both connectivity and node feature information as part of an edge reconstruction task. Good GAE performance would indicate good node embeddings and a more stable and homogeneous market as homogeneity in edge connectivity helps reconstruction performance. Assuming the GAE is well trained, it should be able to group companies of similar returns together. If, once trained, the GAE performs poorly in reconstructing structures of financial graphs of future time steps, we argue this indicates a shift in the way connections are formed in the future graph. 
We interpret this shift as an increased instability in the financial network. Indeed, if the GAE performs worse, then the future graph has heterogeneous connections between companies of different returns and different sectors. A worse reconstruction performance then reflects the definition of a non-homogeneous, unstable graph from literature \cite{Kukreti2020,Onnela2004,Xue2022}. %GAEs have been shown to learn good embeddings useful in downstream clustering tasks \citep{Turner2021}. However, no methodologies have so far exploited the change in accuracy between successive graphs constructed from correlation matrices to explain market instability. We corroborate this understanding by measuring correlation between auto-encoder edge reconstruction performance and market volatility. Finally, to assess the usefulness of using GAEs to capture changes in the structure of a financial market graph, we use the auto-encoder performance to predict future market instability.   % as  with evolving, multitiered structures of interconnections between firms leaving them exposed to increased instability and tail-risk events. even as teir diversification allows them to weather lower-intensity events. 

\section{Problem setting and methodology}

\subsection{Problem setting}
\label{sec:problem}
Consider a financial network represented by a graph $\mathcal{G}=\{\mathcal{E},\mathcal{V}\}$, with nodes representing firms and edges the similarities or correlations between firm prices. In addition, consider a matrix $\mathbf{X}$ where the $u$-th column of $\mathbf{X}^T$, $\mathbf{x}_u$, is a vector of features %$\mathbf{x}_u,~\forall u \in \mathcal{V}$, 
that characterise the firm $u$ (e.g., its stock prices at given time intervals). Our goal is to develop a measure of the instability of the financial market given $\mathcal{G}$ and $\mathbf{X}$. 

Recent works have analysed market instability by investigating the connectivity of $\mathcal{G}$ alone \citep{Onnela2004,Samal2021}. To take into account further information in $\mathbf{X}$, we utilise the GAE proposed in \citep{Kipf2016} which learns vector representations of nodes using both $\mathcal{G}$ and $\mathbf{X}$. The auto-encoder is trained on an edge reconstruction task described in Sec. \ref{subsec:proposed-methodology} where we use a random subset of edges for training and the remaining edges for testing. A good performance on this task would indicate a more homogeneous pattern of edge formation across the graph. 

With this understanding, we frame the problem of measuring market instability as one that quantifies heterogeneity in connectivity patterns between different areas of $\mathcal{G}$. As shown in \citet{Kukreti2020}, homogeneous connectivity patterns are usually observed in low volatility periods where companies in related sectors are often exposed to similar fluctuations and overall more connected to each other than to companies outside their industries. On the other hand, more volatile markets are characterised by highly perturbed correlation structures leading to areas of the graph with highly different connectivity patterns. 
%Homogeneity of connection can be see as the ability of a model trained on an edge reconstruction task over a subset A of edgeset E to reconstruct edges in the edgeset outside A. 
We therefore treat the generalisation ability of the edge reconstruction task above as a proxy for market instability. 
A high reconstruction performance on average with a randomly chosen split of training and test sets would indicate more stable market, while a low performance suggests the opposite due to heterogeneous connectivity patterns across the graph.
\subsection{Data}
\label{sec:data}
We provide a practical context of the problem described above.
Our data set consists of 6 years of stock  market price data  for stocks in the S\&P 500 at the minute-level frequency from 2015 to 2021\footnote{The dataset is obtained from the EOD intra-day data API: \url{https://eodhistoricaldata.com/financial-apis/intra-day-historical-data-api/}. We had to drop some companies as they had no available data during the whole period leaving us with 401 companies.}. 
From these price movements which we define as $p(n,t)$ where where the $nt$-th entry is the price for company $n$ at $t$, we calculate a log return matrix $\mathbf{X} \in \mathbb{R}^{N \times T}$ of $N$ companies over $T$ time periods at frequency $\Delta t$ :  

\begin{equation}
\label{eqn:returns}
r(t, n) =  \text{log} (p(t, n)) - \text{log} (p(t - \Delta t, n))
\end{equation}
% These returns together form a matrix X of returns . 
% \begin{equation}
% \label{eqn:returns_matrix}
% \forall(u) \in V  X_{u} =  r(t, u)
% \end{equation}
Based on this return matrix, we calculate a Pearson correlation matrix based on a rolling window of length $S$ days. This correlation matrix measures for all stock return pairs the covariance normalised by the square root of the product of the variances. It represents the strength of linear relationships between all the stock pairs. Given a return matrix of of $N$ companies over $T$, this gives us $T-S$ correlation matrices in total. 
%Equation \ref{eqn:correlation-matrix} describes C, the rolling Pearson correlation coefficient between two series of returns $\mathbf{x}$ and $\mathbf{y}$ with $\overline{\mathbf{x}}$ and $\overline{\mathbf{y}}$ their respective averages over period R: 
%\begin{equation}
%    \label{eqn:correlation-matrix}
%    C = \frac{{}\sum_{i=t-R}^{t} (\mathbf{x}_i - \overline{\mathbf{x}})(\mathbf{y}_i - \overline{\mathbf{y}})}
%{\sqrt{\sum_{i=t-R}^{t} (\mathbf{x}_i - \overline{\mathbf{x}})^2(\mathbf{y}_i - \overline{\mathbf{y}})^2}}
%\end{equation}
% Using the methodology for adjacency matrix generation from a correlation matrix in \citep{Kukreti}, we apply a thresholding to the correlation. The threshold we set is 0.7, 
Following \citet{Kukreti2020}, we define an adjacency matrix $\mathbf{A}$ as $\mathbf{A}_{uv} = 1$ if the ($u,v$) return pair has a correlation higher than 0.7 and otherwise we set $\mathbf{A}_{uv}$ to 0. The 0.7 threshold  is selected to only keep strongly connected pairs and discount correlations of insufficient strengths \cite{Caccioli2017}\cite{Kukreti2020}.  

We calculate the return volatility ($\text{RV}$) that is used throughout the following sections as a proxy for instability according to the formula set out in Eq.~(\ref{eqn:RV}). It is a measure of the variance of the returns of the average price returns over $\Delta t$ periods: 
\begin{equation}
    \label{eqn:RV}
    \text{RV}_{t}^{t+\Delta t} = \sum_{t}^{t+\Delta t}(r(t,n)^2)
\end{equation}
Because of the power law distribution of market volatility with orders of magnitude larger swings on high-volatility days, volatility is often measured as the natural logarithm of $\text{RV}$ and termed $\text{log}$-$\text{RV}$.

\begin{figure}[!ht]
    \centering
    \includegraphics[width=1\textwidth]{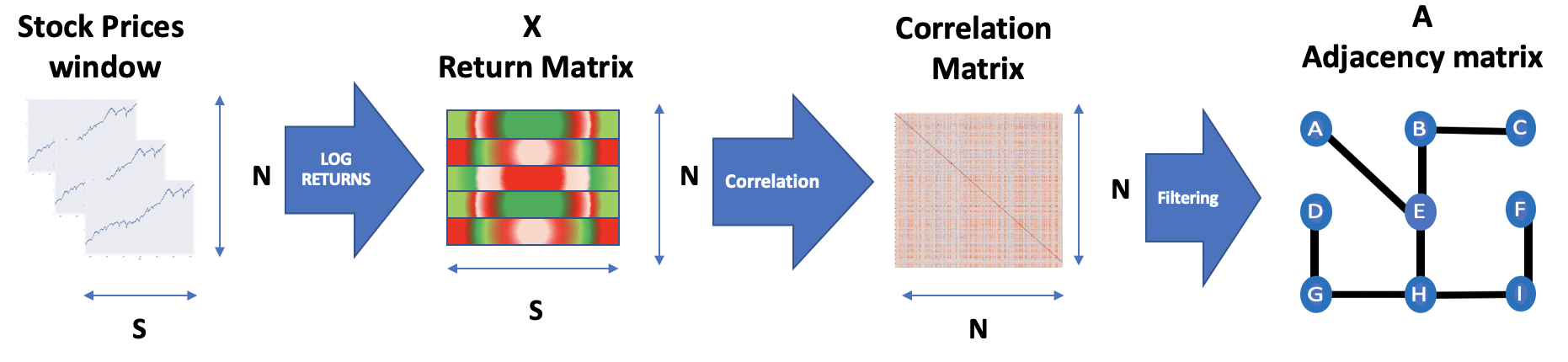}
    \caption{Pipeline generating networks from N returns over a window of time S}
    \label{fig:pipeline_gae}
\end{figure}

We summarise the way we construct the financial network in Fig. (\ref{fig:pipeline_gae}) where a series of stock prices is transformed into a return matrix. From those returns, we calculate the correlation matrix as described above. Lastly, through a threshold applied to that correlation matrix, we obtain a network of connected companies.

\subsection{Proposed methodology}
\label{subsec:proposed-methodology}
%  generalisation of edge prediction model 
Based on the problem setting and data description in the previous two sections, we now formally describe our methodology. 
Given the financial graph $\mathcal{G}$ with adjacency matrix $\mathbf{A}$ and node features $\mathbf{X}$ defined above, we propose to use the GAE \citep{Kipf2016} as the edge reconstruction model whose generalisation ability is used to approximate market instability. 

The GAE is an unsupervised representation learning model with an encoder and a decoder. The encoder is a function defined in Eq.~(\ref{eqn:encoder}). It describes an operation mapping the input $\mathbf{X}$ to the latent embedding $\mathbf{Z}$ via a 2-layer graph convolutional network (GCN) encoder \citep{Kipf}. Once the encoder generates the embedding $\mathbf{Z}$, the decoder in equation Eq.~(\ref{eqn:decoder}) aims to reconstruct the edges present in the initial adjacency matrix. %We are interested in learning the best possible parameters for our encoder so that Z best represents the data. 
The embedding $\mathbf{Z}$ is of “good quality” if the reconstruction $\mathbf{\hat{A}}$ is close to the initial adjacency matrix. The GAE is trained by minimising a binary cross-entropy (BCE) loss defined in Eq.~(\ref{eqn:BCE-Loss}) on training edges $\mathbf{A}_{uv}$:
% that trains our node embedding encoder against a edge reconstruction task. 

%It functions in the graph domain, as shown in \ref{fig:gcn-operations}, akin to a CNN in the image domain : by grouping areas of similar signal together. 
\begin{equation}
\label{eqn:encoder}
\mathbf{Z} = \text{GCN}(\mathbf{A},\mathbf{X}) = \mathbf{\tilde{A}}*\text{ReLU}(\mathbf{\tilde{A}}*\mathbf{X}*\mathbf{W_0})*\mathbf{W_1}
\end{equation}
\begin{equation}
\label{eqn:decoder}
\forall(u,v) \in \mathcal{V}*\mathcal{V}, \mathbf{\hat{A}} = \sigma(\mathbf{z_u}*\mathbf{z_v}) = \frac{1}{(1+e^{-\mathbf{z_u}*\mathbf{z_v}} )}
\end{equation}
\begin{equation}
\label{eqn:BCE-Loss}
(\mathbf{A}_{uv},  \mathbf{\hat{A}}_{uv}) = -\mathbf{A}_{uv}*log( \mathbf{\hat{A}}_{uv}) -(1-\mathbf{A}_{uv})*log(1- \mathbf{\hat{A}}_{uv})
\end{equation}

With the following definitions : 
\begin{itemize}
    \item $\mathbf{\tilde{A}}= \mathbf{D}^{-\frac{1}{2}}\mathbf{A}\mathbf{D}^{-\frac{1}{2}}  $ is the normalised adjacency matrix of the graph. Where $\mathbf{D}=diag(d)$ is the diagonal matrix of the graph $\mathcal{G}$ where the values on the diagonal correspond to the degree of the node. 
    \item \text{ReLU}($\mathbf{x}$) = max($\mathbf{x}$,0) is the non-linear activation function of the GCN. 
    \item $\mathbf{W_0}$ and $\mathbf{W_1}$ are learnable weight matrices   
\end{itemize}

The end-to-end training is done using gradient descent for which we selected a standard Adam algorithm \citep{Kingma2015}.
%Besides the BCE-Loss, we also measure average precision and the area under the ROC curve as the evaluating measures. 
Fig. \ref{fig:gaefromgala} below summarises how the GAE operates.

\begin{figure}[!ht]
    \centering
    \includegraphics[
    width=0.8\textwidth, %height = 35mm
    ]{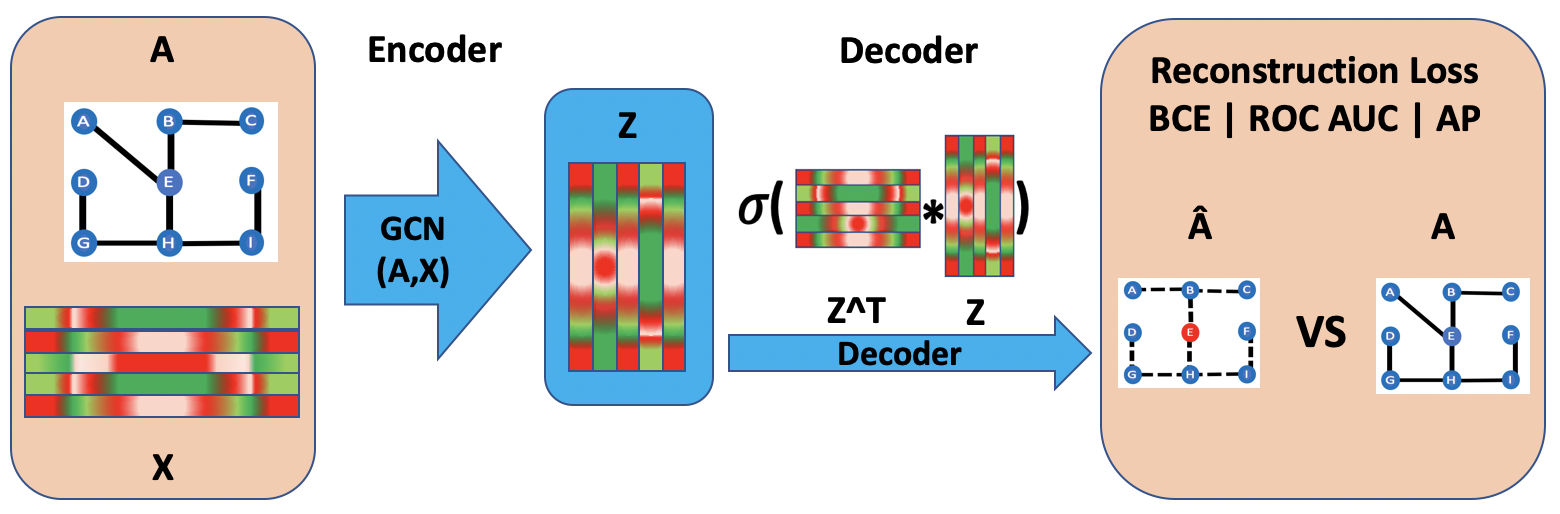}
    \caption{Summary of GAE operations}
    \label{fig:gaefromgala}
\end{figure}

We now describe how we make use of the GAE model to derive a measure of market instability. 
We first extract the returns data $\mathbf{X}$ from a window of length $S = 20$ days. From those returns, we generate the corresponding market graph $\mathcal{G}_t$ with $t$ denoting the last day in the time window using the methodology described in  \ref{sec:data}.  We selected a span of 20 days as this represents one month of trading data, which is long enough to capture relations between companies without too much noise from unstable correlations in shorter windows \cite{Kukreti2020}. Shifting the window by one day, we generate the next day market graph $\mathcal{G}_{t+1}$. 

We trained and validated the  GAE  on $\mathcal{G}_t$,  and test its performance by applying the trained model to reconstruct edges in $\mathcal{G}_{t+1}$, the market graph of next day $t+1$. We assume the GAE is trained on a period of stable market, which means a graph with spatially homogeneous connectivity structures. %this makes the training task on $\mathcal{G}_t$ simple. 
Following the reasoning in Section \ref{sec:problem}, once a GAE is trained, it can reconstruct the adjacency matrix of an unobserved graph in proportion to how spatially homogeneous the unobserved graph is \cite{Kipf2016}. Consequently, any drop of the  GAE testing accuracy on $\mathcal{G}_{t+1}$ can be viewed as an increase in graph heterogeneity. This increased heterogeneity measures how different the edge formation mechanism in $\mathcal{G}_{t+1}$ is different from $\mathcal{G}_{t}$. Previous results in the literature, such as \citet{Zhao2016} suggest that the change in market network homogeneity, measured by edit distance of edges between two successive time periods, granger-causes increases in market volatility. Following  that approach, we interpret the difference in edge formation mechanism between $\mathcal{G}_t$ and $\mathcal{G}_{t+1}$ as resulting from shocks to investor opinion which push the market towards a more volatile state \cite{Lim,Preis2012}. These shocks also form the basis of market instability, and thus we expect this change in network homogeneity between day $t$ and day $t+1$ to correlate with increased volatility on day $t+2$. The reconstruction accuracy measure which we use in our paper to signify the next day edge reconstruction accuracy for any given test graph $\mathcal{G}_{t+1}$ is area under receiver operating characteristic curve ($\mathbf{AUROC}_{t+1}$). We recognise that this approach leads to leakage in the quality of the reconstruction between the training period covered by $\mathcal{G}_t$ and the testing period covered by $\mathcal{G}_{t+1}$ and that this is an issue in direct predictive settings. However, here the main task task is forecasting volatility not graph prediction. As such, we use the training of the GAE only as a synthetic  objective to produce $\mathbf{AUROC}_{t+1}$ which serves only as the input for a downstream volatility measurement. As such, the data leakage is only present in the synthetic task and as there is no leakage in calculating the volatility measurements, our main task of volatility forecasting does not suffer from leakage.
The process is summarised in Fig. \ref{fig:process}  below, where the shift between data at time point $t$ and $t+1$ is shown as a shift of  the two time series $\mathbf{X}_t$ and $\mathbf{X}_{t+1}$, with the former being  used to train the GAE and the latter for testing.
%to test it, generating $\mathbf{AUROC}_{t+1}$. 

\begin{figure}[!ht]
\centering
\includegraphics[width=0.9\textwidth]{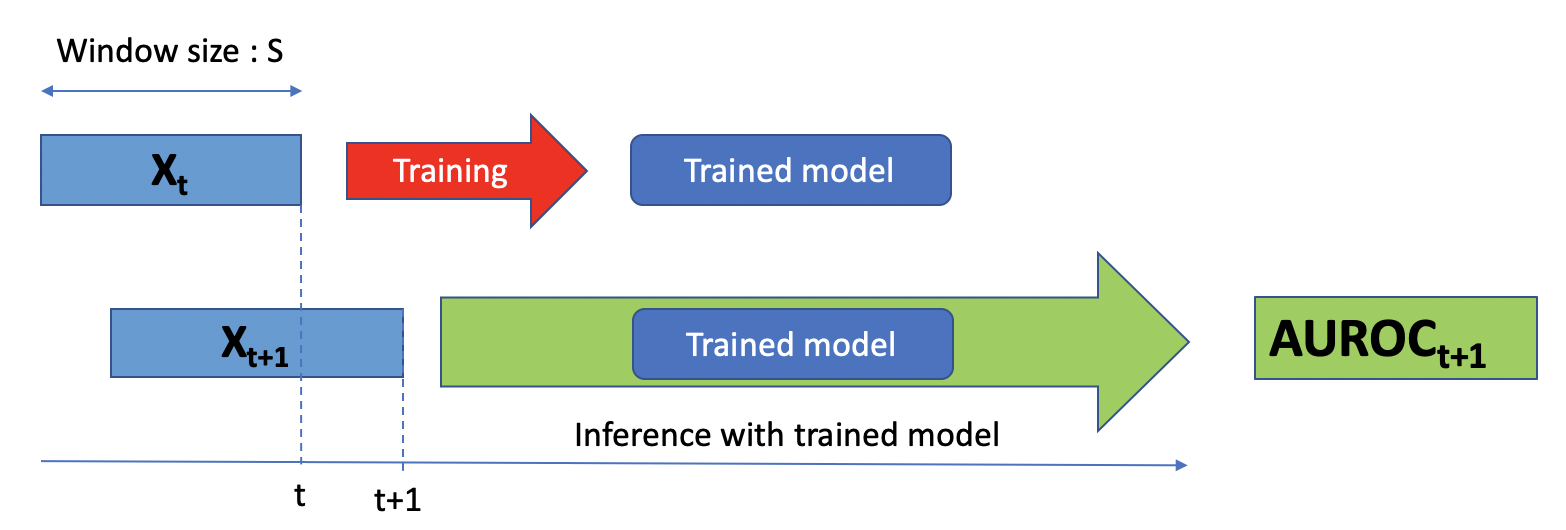}
    \caption{Process of generating $\mathbf{AUROC}_{t+1}$ from model trained at $t$}
    \label{fig:process}
\end{figure}

To asses the performance of the proposed measure, we will first look at correlations between $\mathbf{AUROC}_{t+1}$ (the next day GAE reconstruction) and $\text{RV}_{t+1}^{t+2}$ (the following day's return volatility). This comparison is done by observing the variation of the time series, observing the KDE plots and the Spearman rank correlation of $\mathbf{AUROC}_{t+1}$ and $\text{RV}_{t+1}^{t+2}$. We chose the Spearman rank correlation as it is less vulnerable to outliers. High correlation between $\mathbf{AUROC}_{t+1}$ and $\text{RV}_{t+1}^{t+2}$ would indicate that changes in $\mathcal{G}_{t+1}$'s homogeneity compared with $\mathcal{G}_{t}$  are in sync with market instability. 

As a second step, we then test the usefulness of $\mathbf{AUROC}_{t+1}$ for the forecasting of $\text{log}(\text{RV})$ at the following ${t+2}$ timestep ($\text{log}(\text{RV}_{t}^{t+2})$). We perform this forecasting with the HAR model for log-RV forecasting by \citet{Corsi2009}. The $\text{HAR}$ is a strong and industry recognised benchmark model  for volatility forecasting \citet{Lim}.  It uses auto-regressive components of lower frequency past volatility to predict future volatility. Lower frequency volatility measurements are weekly and monthly volatility. Eq. \ref{eqn:HAR-RV-equation} shows how the the $\text{HAR}$($\text{RV}_{t+2}$) operates, it learns a parameter $\beta_n$ for each previous volatility measure from the past time periods. We look at an 1-hour frequency of forecasting as it is a frequency of interest for industry practitioners \cite{Andersen2006, Onnela2004,Lim}.  We calculate the $R^2$ of this estimation for a linear regression, a XGBoost tree and an MLP as well as the $R^2$ of the estimation to which we add the $\mathbf{AUROC}_{t+1}$ as in Eq.  \ref{eqn:HAR-RV-equation-modif}. An increase  in the  $R^2$ of the forecasting model with the GAE-based measure added would suggest that the GAE encodes useful information for out-of-sample stability forecasting. 
\begin{equation} \label{eqn:HAR-RV-equation}
    \text{log}(\text{RV}_{t+1}^{t+2}) = HAR(\text{log}(\text{RV}_{t}^{t+1})) \sim \beta_1*\text{log}(\text{RV}_{t}^{t+1})+\beta_2*\text{log}(\text{RV}_{t-6}^{t+1})
\end{equation}
\begin{equation} \label{eqn:HAR-RV-equation-modif}
    \text{log}(\text{RV}_{t+1}^{t+2})
\sim HAR(\text{log}(\text{RV}_{t}^{t+1})) + \beta_3*\mathbf{AUROC}_{t+1}
\end{equation}

%Fig. \ref{fig:homogeneity-recon} summarises our method to generate a timeseries of connectivity homogeneity change. 
%\begin{figure}[!ht]
%\includegraphics[width=\textwidth]{images/time_series/homogeneity_extraction.png}
%    \caption{Homogeneity reconstruction}
%    \label{fig:homogeneity-recon}
%\end{figure}
%Therefore, the out-of-sample reconstruction error serves as a good proxy for how much instability     (continue with link to instability and correlation with RV)

\section{Results}
 
\subsection{Correlation between \ensuremath{\mathbf{AUROC}} t+1 and volatility}
\label{subsec:Results1}
%In this section we will first look in figures \ref{fig:timeseries-gae}  and \ref{fig:kde-gae}, at the correlation between the $\mathbf{AUROC}_{t+1}$ and $RV_{t}^{t+1}$ as a preliminary measure for how well our proposed measure reflects market instability. 

%\begin{figure}
%\begin{subfigure}{1.05\textwidth}
%  %\centering
%  \hspace*{-4cm}\includegraphics[width=\linewidth]{images/cleanimages/auc (1).png}
%  
%  \caption{1a}
%  \label{fig:sfig1}
%\end{subfigure}%
%\begin{subfigure}{.30\textwidth}
%  %\centering
%  \hspace*{-4cm}\includegraphics[width=\linewidth]{images/cleanimages/kdeplotsnew (3).png}
%  \hspace*{-4cm}\caption{1b}
%  \label{fig:sfig2}
%\end{subfigure}
%\caption{plots of....}
%\label{fig:fig}
%\end{figure}

\begin{figure}[!ht]
\includegraphics[width=\textwidth]{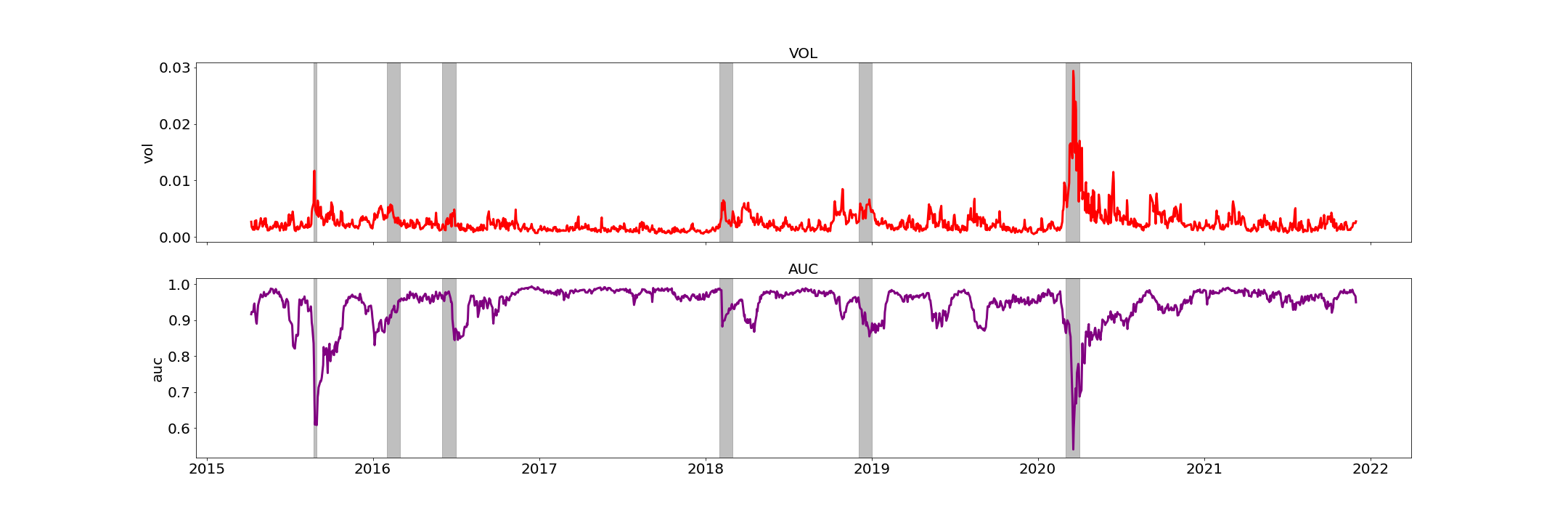}
    \caption{Time series of volatility vs $\mathbf{AUROC}_{t+1}$}
    \label{fig:timeseries-gae}
\end{figure}
When comparing the $\mathbf{AUROC}_{t+1}$ and market volatility, we remark that $\mathbf{AUROC}_{t+1}$ appears to co-move with volatility. $\mathbf{AUROC}_{t+1}$ is high in periods of relative market stability where volatility is lower and sharply fall during downturns and high-volatility markets where volatility is high.
%. Reconstruction loss behaves in the opposite way : it is high when markets are volatile and low when they are stable. 
As GAE model performance is linked to the homogeneity of edge formation across a graph, the model shows lower performance during high-volatility periods. This suggests a more volatile market at $t+1$ correlates to heterogeneous edge formation in the market graph $\mathcal{G}_{t+1}$ which the GAE cannot learn as easily as those in more stable settings confirming the expectation laid out in the literature and our problem setting.  

The six grey bands highlight periods of market instability, either due to so called  `flash crashes' as in late 2015 or those due to fundamental shifts in market stability such as the 2020 pandemic stock  market collapse. These further allow us to remark that the $\mathbf{AUROC}_{t+1}$ seem to dip slightly before a volatility upswing,  suggesting a useful application of these measures in forecasting, which is further investigated in the regression analysis results presented in Sec. \ref{subsec:Results2}.  %Drawdowns and volatility appear strongly correlated, in line with the expectations that stock  markets may grow at a steady non-volatile pace but usually collapse at a rapid pace.

%These observations align with hypothesis 1 that the graph auto-encoder's performance captures elements of market stability due to the difficulty in embedding a graph signal and connectivity that are more unstable in times of high volatility. 
%Moreover, the reconstruction measures in \ref{fig:timeseries-gae}, especially the area under curve and average precision 

\begin{figure}[!ht]
    \centering
    \includegraphics[width = 35mm,height = 35mm]{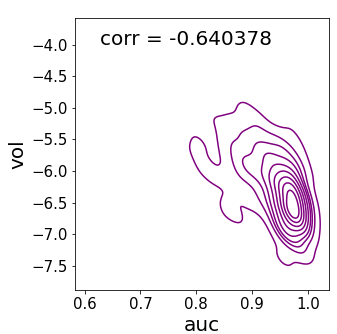}
    \caption{KDE of $\mathbf{AUROC}_{t+1}$ compared to volatility}
    \label{fig:kde-gae}
\end{figure}
The plots above show the kernel density distribution and Spearman rank correlation between log volatility and the $\mathbf{AUROC}_{t+1}$. 
%They reinforce the observations from the previous time-series plots linking market state and the encoder-decoder results. 
The relationship between log volatility and the $\mathbf{AUROC}_{t+1}$ and the correlation in Fig. \ref{fig:kde-gae} displays  an inverse relation between market stability and the GAE's performance. This provides evidence in support of the expectation that the GAE produces a good measure for market instability as it encodes graph homogeneity, and drops in $\mathbf{AUROC}_{t+1}$ are correlated with increased volatility.  

%\begin{table}[!ht]
%\begin{tabular}{llll}
%\toprule
%Model &  MSE with  GAE-ARR & MSE without GAE-ARR &  P-value of difference \\
%\midrule
%Linear & $\mathbf{0.360279^*}$  & 0.107617 &  0.00 \\
%Tree & $\mathbf{0.357711^*}$ &	0.310005 & 0.043 \\
%MLP   & $\mathbf{0.380436^*}$ &	0.317938 & 0.003 \\
%\bottomrule
%Note : * signifies p-value<0.05
%\end{tabular}
%\caption{Results of Log-RV forecasting at the 1 hour frequency}
%
%\label{tab:my-table}
%\end{table}

\subsection{Next-day volatility forecasting}
\label{subsec:Results2}

In this  section, we move on from looking at how well the $\mathbf{AUROC}_{t+1}$ represents market instability. We shift our focus to looking at the usefulness of the $\mathbf{AUROC}_{t+1}$ to predict  $\text{log}(\text{RV}_{t}^{t+1})$ the following day's volatility.  
%power of a volatility forecasting model with the addition of out-of-sample GAE reconstruction accuracy and show that it is useful for measuring next time step volatility.
\begin{table}[!ht]
\begin{center}
\begin{tabular}{||c| c | c | c||} 
 \hline
 Model &  $R^2$ with  $\mathbf{AUROC}_{t+1}$ & $R^2$ without $\mathbf{AUROC}_{t+1}$ &  $p$-value of difference  \\ [0.5ex] 
 \hline\hline
 Linear & $\mathbf{0.360^*}$  & 0.107 &  0.00 \\ 
 \hline
 Tree & $\mathbf{0.357^*}$ &	0.310 & 0.043 \\
 \hline
 MLP   & $\mathbf{0.380^*}$ &	0.317 & 0.003 \\ [1ex] 
 \hline
\end{tabular}
%\begin{tablenotes}
          %\footnotesize   %% If you want them smaller like foot notes
          \item Note : * and bold font indicate $p$-value < 0.05.
%\end{tablenotes}
\end{center}

\caption{Results of $\text{log}$-$\text{RV}$ forecasting at the 1-hour frequency}
\label{tab:my-table}
\end{table}

%\begin{table}[!ht]
%\begin{tabular}{|llll|}
%%\setlength\tabcolsep{4.5pt}
%\toprule
%Model &  $R^2$ with  $\mathbf{AUROC}_{t+1}$ & $R^2$ without $\mathbf{AUROC}_{t+1}$ &  $p$-value of difference \\
%\midrule
%Linear & $\mathbf{0.360^*}$  & 0.107 &  0.00 \\
%Tree & $\mathbf{0.357^*}$ &	0.310 & 0.043 \\
%MLP   & $\mathbf{0.380^*}$ &	0.317 & 0.003 \\
%\bottomrule
%Note : * indicates $p$-value < 0.05
%\end{tabular}
%\caption{Results of Log-RV forecasting at the 1-hour frequency}
%
%\label{tab:my-table}
%\end{table}

Table \ref{tab:my-table} presents the results of the $\text{log}$-$\text{RV}$ forecasting task at time period $t+2$ defined in the methods by Eq. \ref{eqn:HAR-RV-equation} and Eq. \ref{eqn:HAR-RV-equation-modif}. The table compares the $R^2$ with and without the $\mathbf{AUROC}_{t+1}$. The $p$-value displayed on the  rightmost column corresponds to a bootstrapped estimate of the significance of the difference between the MSE of the prediction in a one-sided statistical significance test checking whether the $R^2$ with the added $\mathbf{AUROC}_{t+1}$ is larger than the $R^2$ without. The $H_0$ null hypothesis is that the $R^2$ with $\mathbf{AUROC}_{t+1}$ is not significantly larger than the one without while  the $H_1$ is that there is a statistically significant increase in $R^2$. The results show a statistically significant positive effect of the $\mathbf{AUROC}_{t+1}$ on the forecasting of hourly volatility. This result suggests that the $\mathbf{AUROC}_{t+1}$ of the GAE provides a signal that is useful in forecasting $t+2$ volatility.

\section{Discussion and Conclusion}
Our research investigated the link between the encoding ability of an unsupervised node-embedding model and market instability. The results above support our hypothesis that the the edge reconstruction  accuracy of a GAE would serve as a good proxy to detect changes in the homogeneity of connectivity across a financial graph and thus serve as a good proxy for market volatility. The GAE's encoding of market returns and connectivity at time step $t+1$ is correlated with market instability at the same time step as shown in Section \ref{subsec:Results1} and useful in forecasting out-of-sample volatility as demonstrated in the regression analysis of Section \ref{subsec:Results2}.  The novelty of our contribution lies in underlining the importance of looking at node features as well as the adjacency matrix when deriving stability indicators from financial networks. Moreover,  we extended the literature on the use of graph neural networks in finance by successfully applying an unsupervised methodology to generating good-quality  latent node representations which encode financial states. 

Our approach is flexible and can be applied to other financial networks used in the literature to describe inter firm relations. Thus we aim to apply this approach to graphs constructed from news corpuses  \citep{Heiberger2014} \cite{Wan2021}, supply chains \citep{Matsunaga2019} or knowledge graphs \citep{Hamilton2020}. These other types of graphs capture qualitatively different connections than correlations. That allows them to go beyond the statistical correlations we use to build the graph in this paper. This would allow us to learn instability-predictive features derived from other types of co-movement relations, and in turn refine the volatility forecasting. This same flexibility would allow us to study other edge formation structures which have seen intense interest from the standpoint of propagation of systemic risk such as inter-banking markets \citep{Acemoglu2015} or national stock markets indices. This methodology requires data on price movements of all considered companies for at least 20 days of price returns in order to generate a network. Thus one limitation of the proposed approach is that it cannot deal well with emergent companies with incomplete price data as those new firms price information could not be included in the network formation step.
%This link between homogeneity of different areas and overall economic indicators of stability in the system aligns with results from \citet{Xue2022} which describe how the homogeneity of city transportation graphs is connected to economic returns city-wide. As such, we believe the approach we describe in our work has applications in other fields in estimating other economic indicators.    

%The results in Fig. \ref{fig:timeseries-gae} align with our hypothesis that more volatile markets have a distribution of connectivity and returns which makes their homogeneity harder to reconstruct from latent embedings. 

%This solidifies the increasing interest in graph signal as well as graph topology from the financial community.  graph homogeneity analysed by the prism of graph signal instead of pure This observation also supports previous findings illustrating the link between graph and return timeseries features \citep{Samal2021} and market stability. 
%Our proposed approach learns learns latent representations correlated with market states from both returns timeseries and graphs. 
%Moreover, in the  results from \ref{fig:kde-gae} and \ref{tab:my-table}, the reconstruction task served as a useful signal in out-of-sample forecasting of financial stability measurements. 
We argue that by showing that homogeneity of graph connections is an important feature in market stability, this work also opens future areas for research. With that in mind, we discuss a set of extensions that address the limitations in our solution. The first proposed extension addresses a limitation of the current implementation related to static embeddings: we are currently learning node embeddings for a single static graph and testing the accuracy of the reconstruction on another static graph. In future work, we would like to explore a dynamic feature to the graph auto-encoder as in DynGAE \citep{Mahdavi}. Another extension stems from the agreement between our unsupervised method using graph features and previously observed good performance of non-graph auto-encoder models in finance which attempted to reconstruct only returns  \citet{Lim}. We propose extending the current implementation to reconstruct both adjacency and returns such as in GALA architecture \citep{Park}.

\small{
\bibliography{doubletap}}

\end{document}